\documentstyle[12pt,astrobib,aas2pp4]{article}
\lefthead{REMBGES ET AL.}  
\righthead{AN APPROXIMATION FOR THE rp-PROCESS}
\newcounter{savefg}
\newcommand{\alphfg}{\setcounter{savefg}{\value{figure}}%
\stepcounter{savefg}\setcounter{figure}{0}%
\renewcommand{\thefigure}{\mbox{\arabic{savefg}\alph{figure}}}}
\newcommand{\resetfg}{\setcounter{figure}{\value{savefg}}%
\renewcommand{\thefigure}{\arabic{figure}}}
\begin{document}
\title{AN APPROXIMATION FOR THE rp-PROCESS}

\author{Felix Rembges, Christian Freiburghaus,
        Thomas Rauscher, Friedrich-Karl Thielemann} 

\affil{Departement f\"ur Physik und Astronomie, Universit\"at Basel,
CH-4056 Basel, Switzerland; http://quasar.physik.unibas.ch/}

\author{Hendrik Schatz, Michael Wiescher}

\affil{Department of Physics, University of Notre Dame, Notre Dame,
IN 46556; hschatz@sulfur.helios.nd.edu, wiescher@nd.edu}

\authoraddr{Felix Rembges\\
         Departement f\"ur Physik and Astronomie\\
         Klingelbergstr. 82\\
         CH-4056 Basel\\
         SCHWEIZ\\
         I: rembges@quasar.physik.unibas.ch}

\begin{abstract}
\noindent
{\normalsize
Hot (explosive) hydrogen burning or
the Rapid Proton Capture Process (rp-process) occurs in a number of
astrophysical environments. Novae and X-ray bursts are the most prominent
ones, but accretion disks around black holes and other sites are 
candidates as well.
The expensive and often multidimensional
hydro calculations for such events require an accurate prediction of
the thermonuclear energy generation, while avoiding full nucleosynthesis
network calculations. 

In the present investigation we present an approximation scheme which leads
to an accuracy of more than 15 per cent for the energy generation in hot 
hydrogen
burning from $10^8$--$1.5\times 10^9$ K, which covers the whole range of
all presently known astrophysical sites. It is based on the concept 
of slowly varying hydrogen and helium abundances and assumes a kind of
local steady flow by requiring that all reactions entering and leaving a 
nucleus add up to a zero flux. This scheme can adapt itself automatically
and covers situations at low temperatures, characterized
by a steady flow of reactions, as well as high temperature regimes where a
$(p,\gamma)$--$(\gamma,p)$-equilibrium is established, while 
$\beta^{+}$-decays or
$(\alpha,p)$-reactions feed the population of the next isotonic line
of nuclei.

In addition to a gain of a factor of 15 in computational speed over a
full network calculation, and an energy generation accurate to more than
15 per cent, this scheme also allows to predict correctly individual 
isotopic abundances.
Thus, it delivers all features of a full network at a highly
reduced cost and can easily be implemented in hydro calculations.

\vspace{0.5 cm}
\noindent
{\em Subject headings\/}:
nuclear reactions, nucleosynthesis, abundances --- stars: novae ---
X-rays: bursts
\/}
\end{abstract}

%\keywords{
%nuclear reactions, nucleosynthesis, abundances --- stars: novae ---
%X-rays: bursts}
\vfill

\section{INTRODUCTION}
Close binary stellar systems can exchange mass, when at least one of the
stars fills its Roche Lobe. 
After a critical mass $\Delta M$ of unburned 
transferred matter is accumulated on the surface of the 
accreting star, ignition sets in, typically under degenerate conditions 
when the accreting object is a white dwarf or neutron star. Degenerate 
conditions for which the pressure is not a function of temperature,
prevent temperature adjustment via pressure increase and expansion
and cause a thermonuclear runaway and explosive burning. 
For white dwarfs, a layer of $10^{-5}$--$10^{-4}$ M$_\odot$ forms, before
pycnonuclear ignition of hydrogen burning sets in
(e.g. \shortciteNP{sugimoto80}; \shortciteNP{starrfield93}; 
\shortciteNP{coc95} ) and 
causes
a nova event (explosive H-burning on white dwarfs) with maximum 
temperatures of $\approx$(2--3)$\times 10^8$ K and a total energy release of 
$10^{46}$--$10^{47}$ ergs.
The burning takes 100--1000 s before the partially burned hydrogen
envelope is ejected.

X-ray bursts (for an observational overview see \citeANP{lewin93} 1993)
involve accreting neutron stars with an unburned, hydrogen-rich surface layer.
The critical size of the hydrogen layer before ignition is as small as 
$10^{-12}$ M$_\odot$.
Temperatures of the order (1-2)$\times 10^9$ K and densities 
$\rho$$\approx$$10^{6}$--$10^{7}$ g cm$^{-3}$ are attained (see e.g.
\shortciteNP{wallace81}; \shortciteANP{ayasli82} 1982; 
\citeANP{hanawa83} 1983;
\citeANP{woosley86} 1986; \shortciteANP{taam93} 1993; 
\citeANP{taam96} 1996).
This explosive burning with rise times of about 1--10 s leads to 
the release of $10^{39}$--$10^{40}$ ergs. Many of the observed
features and general characteristics are understood, however, there is
still a lack of a quantitative understanding of the detailed observational
data. Another aspect is that the explosion energies are smaller than the
gravitational binding energy of the accreted hydrogen envelope. An evenly
distributed explosion energy would not unbind and eject matter. It
remains to be seen whether an interesting amount of matter can escape
the neutron star. Super-Eddington X-ray bursts (\shortciteNP{taam96}) are
the best candidates for this behavior.

A description of hot 
(explosive) hydrogen burning has been given by \shortciteANP{arnould80} (1980),
\shortciteANP{wallace81} (1981),
\shortciteANP{ayasli82} (1982), \shortciteANP{hanawa83} (1983), 
\shortciteANP{wiescher86} (1986), 
\shortciteANP{wormer94} (1994), \shortciteANP{nuc_phys_thie} (1994)
[see also \shortciteANP{biehle91} (1991),  \shortciteANP{biehle94} (1994), 
\citeANP{cannon92}
 (1992), and 
\shortciteANP{cannon93} (1993)
for Thorne-Zytkow objects, but see the recent results by \citeANP{fryer96}
(1996) which puts doubts on their existence]. 
The burning is described by proton captures, $\beta^+$-decays, and possibly
alpha-induced reactions on unstable proton-rich nuclei, usually referred to
as the rp-process (Rapid Proton Capture Process). 
Cross sections can
either be obtained from the best available
application of present experimental knowledge, e.g. 
the determination of resonance properties from mirror nuclei
and transfer and/or charge exchange reaction studies, 
or actual cross section measurements,
like for $^{13}$N($p,\gamma)^{14}$O, the first
reaction cross section analysed with a radioactive ion beam facility
(see \shortciteANP{champagne92} 1992 and references therein).
There exist two major motivations in nuclear astrophysics (a), to understand
the reaction flow to a necessary degree, in order to predict the correct
energy generation required for hydrodynamic, astrophysical studies,
and (b), to predict a detailed isotopic composition which helps to understand
the contribution of the process in question to nucleosynthesis in general.
Our main motivation in this paper is (a), to provide a fast energy generation
network as a tool for nova and X-ray burst studies. This might be underlined
by the fact that novae and X-ray bursts seem not to be major contributors
to nucleosynthesis due to the small ejected masses involved or the 
question whether gravitational binding can be overcome at all.
[However, novae can be important for nuclei like $^{7}$Li, $^{15}$N, 
$^{22}$Na, $^{26}$Al, and even Si and S; and super-Eddington X-ray bursts 
(\shortciteANP{taam96} 1996) will be
able to eject some matter, probably containing some light p-process elements.]
It will turn out at the end that our efficient approximation and energy
generation scheme can also be used to predict abundances accurately.
The nucleosynthesis for nuclei above Kr will, however, be discussed in a
second paper (\shortciteANP{schatz96} et al. 1996).

In order to understand the energy generation correctly, we have to be
able to understand the main reaction fluxes. In the past
we performed a series of rp-process studies (\shortciteANP{wiescher86} 1986 
to Ar, and \shortciteANP{wormer94} 1994, extending up to Kr).
The reaction rate predictions were based on resonance and direct capture
contributions for proton-rich nuclei, making use of the most recent 
experimental data. 
Several ($p,\gamma$)-reaction rates below mass A=44 have been recalculated by 
\shortciteANP{herndl95} (1995) in the framework of a shell 
model description for the  level structure of the compound nucleus.
A preliminary analysis of the major aspects was given by
\shortciteANP{nuc_phys_thie} (1994). In the following 
section \ref{reacflow} we will present this in more detail and
discuss the constraints which an approximation scheme
has to fulfill in the whole range of temperatures occurring in explosive
hydrogen burning environments, i.e. $10^8$--$1.5\times 10^9$ K.
The approximation scheme will be presented in section \ref{approxscheme},
its application and comparison with full network calculations in
section \ref{results}, followed by concluding remarks in section \ref{conc}.

\section{REACTION FLOWS IN EXPLOSIVE HYDROGEN BURNING}\label{reacflow}

\begin{figure}[htb]
\epsscale{0.75}
\figurenum{1}
%\plotone{cycle43.ps}
\plotfiddle{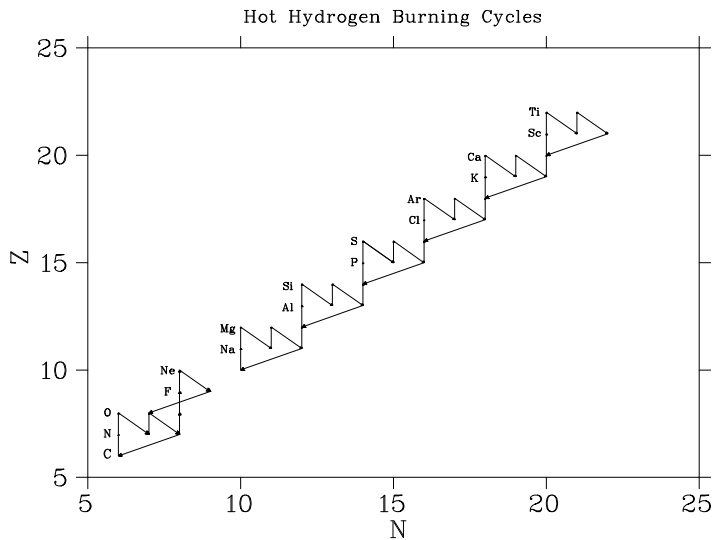}{6 cm}{90}{40}{40}{130}{-15}
\figcaption[cycle43.ps]{The hot hydrogen burning cycles, established at
T$_{9}$=0.3 and $\rho$=$10^{4}$ g cm $^{-3}$.\label{fig cycle43}}
%\caption{The hot hydrogen burning cycles, established at
%T$_{9}$=0.3 and $\rho$=$10^{4}$ g cm $^{-3}$.}
\end{figure}

At low temperatures, the rp-process is dominated by cycles of two
successive proton captures, starting out from an even-even nucleus,
a $\beta^+$-decay, a further proton capture (into an even-$Z$ nucleus), another
$\beta^+$-decay, and a final $(p,\alpha)$-reaction 
close to stability (similar to the hot CNO). The final closing of the cycle
via a $(p,\alpha)$-reaction occurs because proton 
capture Q-values increase with
increasing neutron number $N$ in an isotopic chain for a given $Z$, while
$\alpha$-capture Q-values or -thresholds show a
very weak dependence on $N$. Thus, a cycle closure occurs at the most
proton-rich compound nucleus, which has a lower alpha
than proton-threshold. These are the even-$Z$ and especially well bound
``alpha''-nuclei with isospin T$_z$=$(N-Z)/2$=0,
like $^{16}$O, $^{20}$Ne, $^{24}$Mg, $^{28}$Si, $^{32}$S, $^{36}$Ar, $^{40}$Ca. 
Therefore, $(p,\alpha)$-reactions can operate on odd-$Z$ targets with
T$_z$=+1/2: $^{15}$N, $^{19}$F, $^{23}$Na, $^{27}$Al, $^{31}$P, $^{35}$Cl,
$^{39}$K.
The hydrogen burning cycles are connected at these nuclei, due to  a possible
competition between the  $(p,\gamma)$- and $(p,\alpha)$-reaction. 
An exception to the rule is the Ne isotopic chain, where the transition
happens already for the compound nucleus $^{19}$Ne.
Thus, $^{18}$F($p,\alpha)^{15}$O is a possible reaction, bypassing
the  OF(Ne)-cycle. Therefore, only the extended CN(O)-, the 
NeNa-, MgAl-, SiP-, SCl-cycles etc. exist. This is displayed in 
Figure \ref{fig cycle43}.

\placefigure{fig cycle43}
%\onecolumn
%\clearpage
%\thispagestyle{empty}

The progress of burning towards heavier nuclei depends on the
leakage ratio $(p,\gamma)/(p,\alpha)$ into the next cycle, which makes 
a good experimental determination of these reactions important and
is possible as we deal here with stable targets. On the other hand, the 
excitation energies in the corresponding compound nuclei are of the order 
8.5--12~MeV and
make statistical model approaches also reliable (see e.g. the experiments
by \shortciteANP{iliadis93} 1993, 1994 
and \shortciteANP{ross95} 1995 for $^{31}$P($p,\gamma)$, and 
$^{35}$Cl($p,\gamma)$ and the later discussion).

\begin{figure}[tb]
\figurenum{2}
\epsscale{0.75}
\plotfiddle{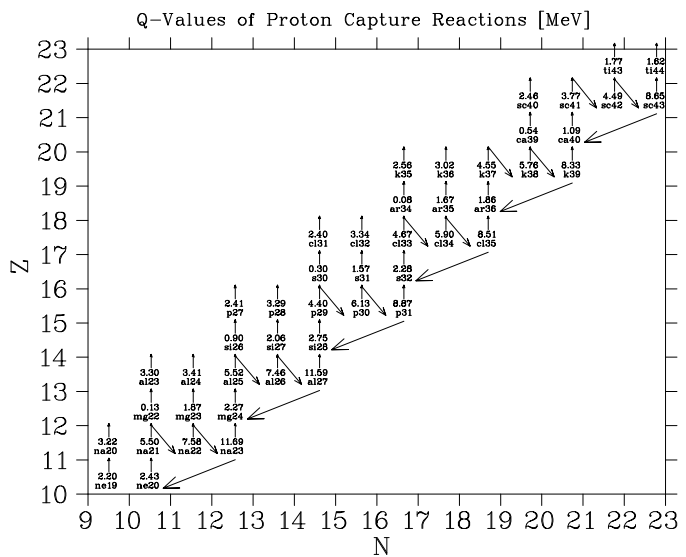}{6 cm}{+90}{40}{40}{160}{-15}
\figcaption[qval1.ps]{Q-values of ($p,\gamma$)-reactions between
 $^{20}$Ne and  $^{44}$Ti.
\label{fig qval1}}
\end{figure}

Increasing temperatures allow to overcome Coulomb barriers and
extend the cycles to more proton-rich nuclei, which permit additional
leakage via proton captures, competing with long $\beta$-decays.
The cycles open first at the last $\beta$-decay 
before the $(p,\alpha)$-reaction via a competing proton capture, i.e.
at the even-$Z$ T$_z$=--1/2 nuclei like $^{23}$Mg, $^{27}$Si, $^{31}$S, 
$^{35}$Ar, and $^{39}$Ca, excluding $^{15}$O (because  $^{16}$F is
particle unstable) and $^{19}$Ne, which is
bypassed by the stronger $^{18}$F($p,\alpha)$-rate. 
These nuclei are only one unit away from stability and
have small decay Q-values and long half-lives.
The situation is illustrated in Figure \ref{fig qval1}, which shows the 
possible
break-out points and the Q-values of the proton capture reactions.
Since the reaction proceeds towards the proton drip-line and into an
odd-odd nucleus, the capture Q-values of these $(p,\gamma)$-reactions
are small (generally less than 2 MeV). This implies a too small density
of excited states for employing statistical model cross sections
(see discussion in the following paragraphs).
At about 3$\times$10$^8$ K essentially all such cycles are open and a
complete rp-pattern of proton captures and $\beta$-decays is established,
with the exception of the extended CNO-cycle. The process time is
determined by the sum of $\beta$-decay and proton capture time scales,
$\tau_\beta$ and $\tau_{p,\gamma}$, along
the rp-path, which does generally not extend more proton-rich than
T$_z$=--1 for even-$Z$ nuclei. The reaction time scales
are dominated by the T$_z$=--1/2 even-$Z$ nuclei close to stability, listed 
above.  

\placefigure{fig qval1}

At 4$\times$10$^8$~K the CNO-cycle breaks open via
$^{15}$O($\alpha,\gamma)^{19}$Ne. $^{14}$O($\alpha,p)^{17}$F can also lead to 
a successful
break-out when followed by a proton capture and $(\alpha,p)$-reaction on
$^{18}$Ne. Due to higher energies, Coulomb barriers can be more
easily overcome by heavier projectiles (like alpha captures on CNO-targets), 
and proton captures can occur further away from stability. However, in all 
cases the
decay of T$_z$=--1 even-$Z$ nuclei (and at high densities the competition 
with $(\alpha,p)$-reactions) is a prominent part
of the reaction path. 
Alpha-reactions are not important for $Z$$>$20,
due to increasing Coulomb barriers and decreasing Q-values. 
For heavier nuclei beyond Ca, the reaction pattern seems 
complicated. However, 
temperatures approaching $10^9$~K cause strong
photodisintegrations for proton capture Q-values smaller than 
25$kT$$\approx$3~MeV,
which is comparable to Q-values found for even-$Z$ T$_z$=--1 nuclei. 
[25-30$kT$$\ge$$Q$ is a rough criterion to find out whether 
photodisintegrations
can counterbalance capture reactions and lead to an equilibrium situation.]
Thus, reactions breaking out of the cycles discussed above (up to the
proton drip line) come into a $(p,\gamma)$--$(\gamma,p)$-equilibrium along
isotonic lines  with equal $N$.  Under such circumstances
the nuclear connection boils down
to the necessary knowledge of nuclear masses and $\beta$-decay half-lives,
similar to the r-process
experiencing an $(n,\gamma)$--$(\gamma,n)$-equilibrium in isotopic chains
(see \citeNP{cowan91}; \shortciteNP{kratz93}; 
\shortciteNP{meyer94}).

\begin{figure}[tb]
\epsscale{0.75}
\figurenum{3}
\plotfiddle{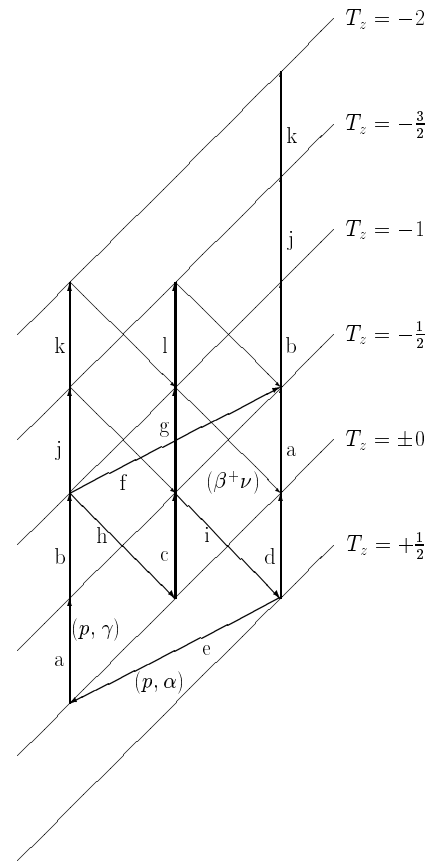}{8 cm}{0}{50}{50}{-150}{-50}
\figcaption[netzw.ps]{Recurrent structure in the rp-process
network.
\label{fig netzw}}
\end{figure}

\placefigure{fig netzw}
%\clearpage
%\thispagestyle{empty}

\begin{figure}[tb]
\epsscale{0.75}
\figurenum{4a}
\plotfiddle{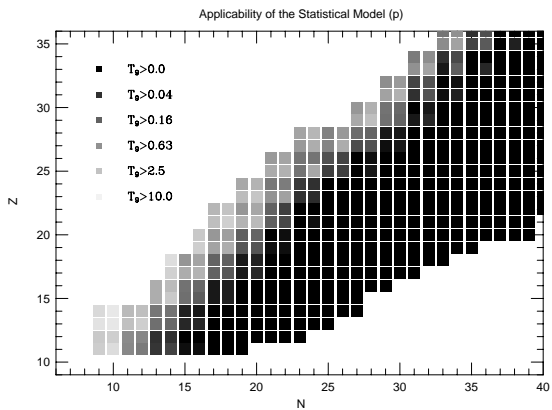}{6 cm}{0}{30}{30}{-120}{0}
\label{fig rp_fig4a}
\figcaption[rp-fig4a.ps]{Applicability of the Statistical Model for
$(p,\gamma)$- and $(p,\alpha)$-reactions.}
\end{figure}

\begin{figure}[tb]
\figurenum{4b}
\epsscale{0.75}
\plotfiddle{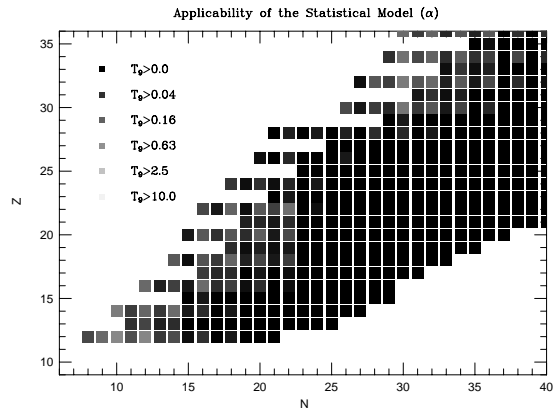}{6 cm}{90}{30}{30}{120}{-0}
\figcaption[rp-fig4b.ps]{Applicability of the Statistical Model for
$(\alpha,\gamma)$- and $(\alpha,p)$-reactions.
\label{fig rp_fig4b}}
\end{figure}

The general situation is explained in Figure \ref{fig netzw}, which contains 
all possible reactions from the original cycles over proton 
captures breaking out of the
cycles, the accompanying beta decays and $(\alpha, p)$-reactions bridging
waiting points with long decay half-lives. By comparing with
Figure \ref{fig qval1}, which also lists the Q-values involved,  
Figures \ref{fig rp_fig4a} and  \ref{fig rp_fig4b} give
an indication for the minimum temperatures required to permit statistical model
calculations for proton and alpha-induced reactions.
The reason is that a high density of compound nuclear states is necessary for
statistical model calculations to become reliable. The energy of the Gamow
peak, where the integrand in a thermonuclear cross section integral
(cross section times Boltzmann distribution) has a maximum, increases with
increasing temperature. Because higher level densities are encountered
at higher excitation energies corresponding to higher temperatures of the
Boltzmann distribution, a critical temperature exists beyond which such
calculations are reliable (for more details see
\citeANP{rauscher96} 1996). We can analyse the expected behavior:
Reactions $j$, $k$, and $l$ of Figure \ref{fig netzw}
have reaction Q-values less than $\approx$3~MeV 
and for conditions close to 10$^9$~K we have 25$kT$$\approx$$Q$ which 
permits a ($p,\gamma)$--$(\gamma,p)$-equilibrium. 
Thus, although the cross sections
are highly uncertain and statistical model approaches are not valid (see
Figure \ref{fig rp_fig4a}), only the Q-values, i.e. masses, 
are needed for these nuclei.
Reactions $b$, $c$, $d$, and $e$ [($p,\gamma)$ and ($p,\alpha)$ reactions] have
Q-values in excess of 5~MeV and in most cases high enough level densities
to ensure the usage of statistical model cross sections for the appropriate
temperature regime. Reaction $f$, an $(\alpha,p)$-reaction, provides also 
a sufficiently high level density to apply statistical model cross sections
(see Figure \ref{fig rp_fig4b}, for $^{18}$Ne($\alpha,p$) see 
\shortciteNP{goerres95b}). 
Reaction $g$ is crucial for the break-out, a statistical
model approach is clearly not permitted, and a 
($p,\gamma)$--$(\gamma,p)$-equilibrium is not valid at the 
lower appropriate temperatures of about
3-4$\times 10^8$~K. Thus, all reactions of type $g$ ask strongly for 
experimental determination, possibly with
radioactive ion beams. Reactions $h$, $i$ and all other connecting 
$\beta$-decays are required, preferably from experiments, otherwise from 
theoretical QRPA predictions.

\alphfg

\placefigure{fig rp_fig4a}
\placefigure{fig rp_fig4b}

Beyond Ca and Ti, this scheme simplifies into one dominated by proton
captures and beta-decays, with the proton captures in a 
($p,\gamma)$--$(\gamma,p)$-equilibrium. 
The proton drip line develops  a zig-zag shape and can 
separate two even-$Z$ proton-stable nuclei on the same 
isotone by an unstable odd-$Z$ nucleus. \shortciteANP{goerres95a} (1995) have 
shown that such gaps can be bridged by two-proton capture
reactions, proceeding through the unstable nucleus, similar to the
$3\alpha$-reaction passing through $^8$Be. This can extend the 
$(p,\gamma)$--$(\gamma,p)$-equilibrium beyond the drip line to 
proton-stable "peninsulas", thus avoiding long beta-decay half-lives.
While alpha-induced reactions are not important for nuclei beyond Ca and Ti,
small  Q-values for $A>56$ can permit $(\gamma,\alpha)$-photo\-disintegrations 
and cause a back flow, decreasing the build-up of heavy nuclei.  

After having analized the reaction patterns in the previous discussion,
we have now to implement these findings into an approximation which
can handle and describe all of the patterns. Considering the hydrogen and
helium abundances as slowly varying, and assuming a kind of local steady
flow by requiring that all reactions entering and leaving a nucleus
add up to a zero flux, we will devise a scheme which can adapt itself to
the whole range of conditions of interest in hot hydrogen burning.
It can automatically cover situations at low temperatures, characterized
by a steady flow through all connected hot CNO-type cycles and branchings
feeding the next cycle, to high temperature regimes where a
$(p,\gamma)$--$(\gamma,p)$-equilibrium is established, while $\beta$-decays or
$(\alpha,p)$-reactions feed the population of the next isotonic line.
The details are discussed in the following section.

\section{APPROXIMATION SCHEME}\label{approxscheme}

\subsection{Reaction Rates and Quasi-Decay Constants}

The thermonuclear reaction rates used in our calculations have
been  discussed in detail by van Wormer et al. 
\citeyear{wormer94} ({\em cf.\/} the appendix therein).\footnote{The 
parameterized  reaction rates can be  found for instance
in the database REACLIB located at 
http://csa5.lbl.gov/ $\tilde{}\,$fchu/astro.html/.} 
The temporal change of the isotopic abundances
$Y_{i}$=$\frac{X_{i}}{A_{i}}$ can be calculated by all depleting and 
producing reactions

\begin{eqnarray}
Y_{i}&=&\sum_{j} {^{i}\alpha_{j}}\,Y_{j}
+\sum_{j,k} {^{i}\alpha_{j,k}}\,Y_{j}\,Y_{k} \nonumber\\
&&+\sum_{j,k,l}  {^{i}\alpha_{j,k,l}} Y_{j}\,Y_{k}\,Y_{l}\, .
\label{eq netpa1}
\end{eqnarray}

\noindent
The first term in (\ref{eq netpa1}) 
includes all one-particle reactions (decays or photodisintegrations)
 of nuclei $j$ producing 
($i$$\not=$$j$) 
or destroying ($i$$=$$j$) nucleus $i$. 
The $\alpha_{j}$ are either decay constants or 
effective temperature dependent decay constants in the case of 
photodisintegrations.
The second sum represents all 
two-particle reactions between nuclei $j,k$ with  
${^{i}\alpha_{j,k}}$=${^{i}c_{j,k}}\,\rho N_{A}{^{i}\langle \sigma 
v\rangle_{j,k}}$. Similarly, the third term represents three particle
processes with 
${^{i}\alpha_{j,k}}$ 
=${^{i}c_{j,k,l}}\rho^{2}$$N_{A}^{2}{^{i}\langle} 
\sigma v\rangle_{j,k,l}$ (\citeANP{fowler67} 1967). The coefficients $c$ include the 
statistical factors for avoiding double counting of reactions of 
identical particles.

\begin{figure}[tbn]
\epsscale{0.75}
\figurenum{5a}
\plotfiddle{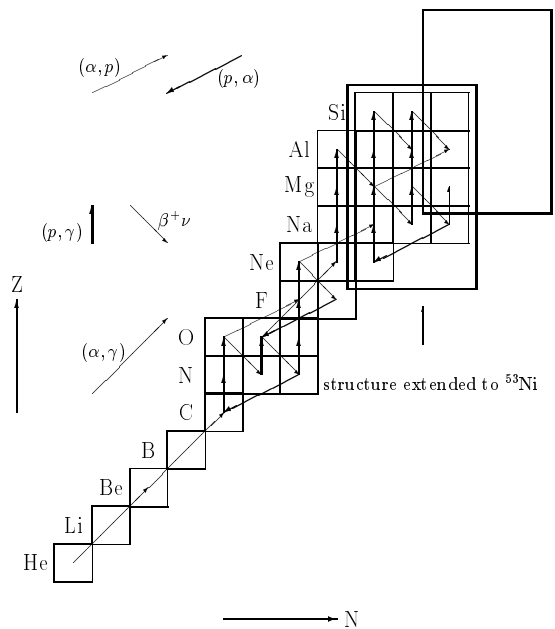}{7 cm}{0}{50}{50}{-130}{-100}
\figcaption[diagpa1.ps]{rp-process network being used in the approximation. 
The structure in the box is eight times repeated up to  $^{50}$Mn.
Bold arrows correspond to large, normal arrows to small  decay constants.
\label{fig diagr1}}
\end{figure}

\begin{figure}[tbn]
\epsscale{0.75}
\figurenum{5b}
\plotfiddle{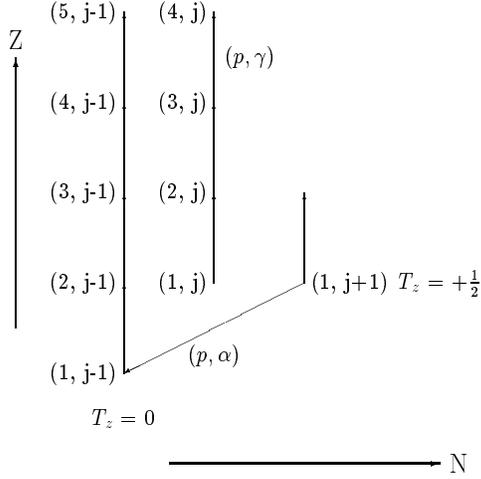}{6 cm}{0}{60}{60}{-200}{-250}
\figcaption[diagpa2.ps]
{The proton capture sequences are connected to each other by ($\alpha,p$)- and
($\beta^{+},\nu$)-breakout reactions respectively. 
\label{fig diagr2}}
\end{figure}

\placefigure{fig diagr1}
\placefigure{fig diagr2}

\resetfg
\alphfg
%\clearpage
%\thispagestyle{empty}

%\clearpage
%\thispagestyle{empty}

van Wormer et al. \citeyear{wormer94}
investigated numerically which reactions must be considered
in the rp-process. Apart from the temperature insensitive
$\beta^{+}$-decays, the only one-particle reactions which 
must be especially considered at high temperatures are photodisintegrations.
The two-body reactions
dominating the rp-process are $(p,\gamma)$-, $(p,\alpha)$-
and, $(\alpha,p)$-reactions. Three body reactions can be neglected
in explosive hydrogen burning with the exception of
the $3\alpha$-reaction which plays a central role
in this context.
% and the the proton captures mentioned in
%section \ref{reacflow}. But the latter become only important at high
%temperatures when ($p,\gamma$)--($\gamma,p)$-equilibria are established.
Taking into account individual reactions, equation
(\ref{eq netpa1}) can be rewritten as

\begin{eqnarray}
\dot{Y}_{i}&=&\sum_{j}\,{^{i}\lambda}^{\beta^{+}}_{j}\,Y_{j}+
        \sum_{j}\,{^{i}\lambda}^{\gamma,p}_{j}\, Y_{j} \nonumber\\
     && +\sum_{j}\,\rho N_{A}{^{i}\langle} \sigma v\rangle_{j,p} Y_{j}Y_{p}
        \nonumber\\
     && +\sum_{j}\,\rho N_{A}
          {^{i}\langle} \sigma v\rangle_{j,\alpha} Y_{j} Y_{\alpha}\; .
\label{eq netpa2}
\end{eqnarray}

\noindent
If the proton and helium abundances do not change greatly during
relevant rp-process time scales, we are able to describe
proton captures and (${\alpha,p}$)-reactions with {\em quasi-decay 
constants}, 
provided that temperature and density do not change greatly during the
process time scale of interest:

\begin{mathletters}
\begin{eqnarray}
 {^{i}\lambda}^{p,\gamma}_{j}&:=& \rho N_{A}\,
{^{i}\langle \sigma v \rangle}_{j,p} Y_{p} =\mbox{const} \\
 {^{i}\lambda}^{p,\alpha}_{j}&:=& \rho N_{A}\,
{^{i}\langle \sigma v \rangle}_{j,p} Y_{p} =\mbox{const} \\
 {^{i}\lambda}^{\alpha,\gamma}_{j}&:=& \rho N_{A}\,
{^{i}\langle \sigma v\rangle}_{j,\alpha} Y_{\alpha} =\mbox{const}\\
 {^{i}\lambda}^{\alpha,p}_{j}&:=& \rho N_{A}\,
{^{i}\langle \sigma v\rangle}_{j,\alpha} Y_{\alpha} =\mbox{const} \, .
\label{eq netpa3}
\end{eqnarray}
\end{mathletters}

%\begin{eqnarray}
%&&\fbox{$ \displaystyle {^{i}\lambda}^{p,\gamma}_{j}:= \rho N_{A}\,
%{^{i}\langle \sigma v \rangle}_{j,p} Y_{p} =const $}\\
%&&\fbox{$ \displaystyle {^{i}\lambda}^{\alpha,p}_{j}:= \rho N_{A}\,
%{^{i}\langle \sigma v\rangle}_{j,\alpha} Y_{\alpha} =const $}\, .
%\label{eq netpa3}
%\end{eqnarray}

\noindent
Thus the system of nonlinear differential equations can be written in a 
linearized form

\begin{equation}
\dot{Y}_{i}=\sum_{j} {^{i}\lambda}_{j}^{x} Y_{j}\; ,
\label{eq netpa4}
\end{equation}

\noindent
where $x$ represents one of
the following reactions: 
$(\gamma,p)$, $(p,\gamma)$, $(\alpha,p)$, $(p,\alpha)$,
 $(\alpha,\gamma)$, $(\gamma,\alpha)$, $(\beta^{+}\nu)$.
When we choose 
${\mathrm Y_{0}}$=$(Y_{1},$ $\dots,Y_{i},$ $\dots,Y_{n})$ to 
represent the initial 
abundance distribution, the system of equations (\ref{eq netpa4})  
can be written as

%\mathversion{bold}
%mathversion{normal}
\begin{equation}
{\mathrm{\dot{Y}\/}}=A\,
{\mathrm {Y\/}}\, , \quad A\, (\mbox{$N\times N$)-matrix}\, .
\end{equation}
%\mathversion{normal}

Such systems of linear differential equations 
with {\em constant} coefficients can be solved
by Jordan-normalizing  matrix $A$. Its 
transformation 

\begin{equation}
J=SAS^{-1}\, ,\,\, \mbox{$S\,
\epsilon\,$GL$_{\mbox{\tiny n}}$}\bigl
(\mbox{\tt{\large{R\hspace{-2.95 mm}R}}}\bigr)
%\epsilon\,$GL$_{\mbox{\tiny n}}$}\bigl({\mathbb R}\bigr)
%\quad \mbox{(= General Linear Group)}  
\end{equation}

\noindent
yields  the $N$-dimensional set of eigenfunctions
 
\begin{equation}
u_{i;j}(t)=e^{\lambda^{m}_{i}\, t}p_{i;j}(t)\, .
\end{equation}

\noindent
Here, ${\lambda}^{m}_{i}$ represents eigenvalue $i$ with degeneracy $m$,
$u_{i;j}$ eigenvector $j$ to eigenvalue $i$ ($1$$\leq$$j\leq$$m$) and
$p_{i;j}(t)$ a polynomial of degree ($j$--$1$).
However, the transformations $u'_{i;j}$=$S\,u_{i;j}$ 
are in general difficult to
survey and to handle. Furthermore, the elements of
$A$ become slowly varying functions when the abundances of hydrogen and
helium  change slightly over a certain process time scale. 
This is for instance the case, when we consider hot hydrogen burning. 
Only numeric simulations will be feasible for these two reasons.

\begin{figure}[t]
\epsscale{0.75}
\figurenum{6a}
\plotfiddle{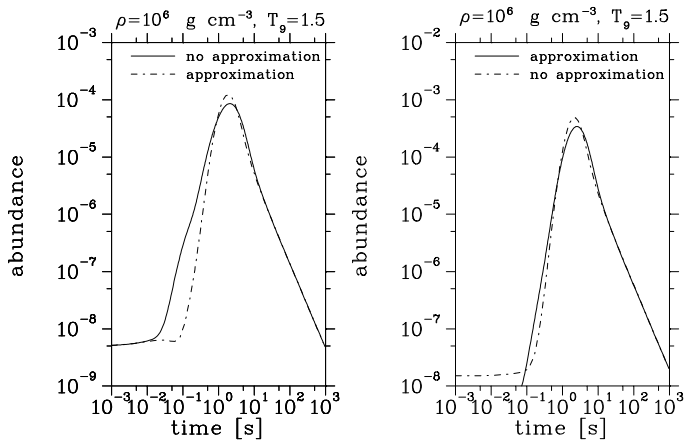}{5 cm}{90}{40}{40}{140}{-50}
\figcaption[ab615p1.ps]{Comparison of the 
approximated abundances of  $^{53}$Ni and  $^{55}$Ni
with results from full network calculations.\label{fig abpa1}}
\end{figure}

\begin{figure}[tbn]
\figurenum{6b}
\epsscale{0.75}
\plotfiddle{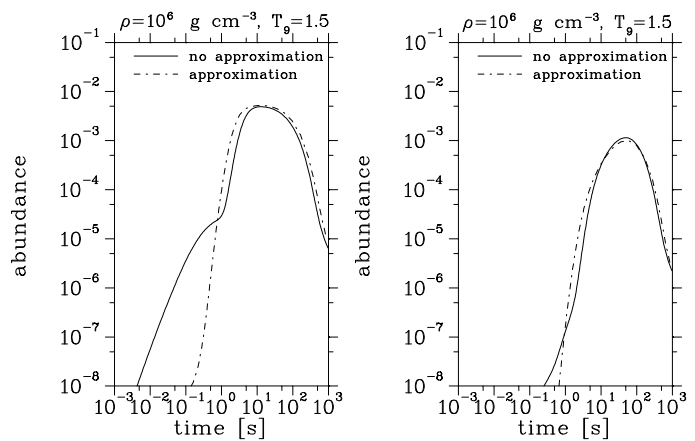}{5 cm}{90}{40}{40}{140}{-50}
\figcaption[ab615p2.ps]{Same as Figure \ref{fig abpa1} but for 
 $^{64}$Ge and  $^{72}$Kr.\label{fig abpa2}}
\resetfg
\end{figure}

\placefigure{fig abpa1}
\placefigure{fig abpa2}
\resetfg
\alphfg
%\clearpage
%\thispagestyle{empty}

%\clearpage
%\thispagestyle{empty}

We will show in the following subsection how certain sequences of a 
thermonuclear reaction network can be regarded as  {\em quasi-nuclei}.
In such sequences a steady flow materializes and the ratios between 
the different abundances are
constant and are only determined by the values of the
quasi-decay constants.
A procedure of this kind provides us with a system  of equations with a 
smaller dimension, which can be solved numerically.
%or: which can be solved very fast by standard numerical techniques.

\subsection{The Approximation}\label{approximation}
In section \ref{reacflow}, we discussed the burning cycles with
the same structure as the hot 
CNO-cycle (see also Figures \ref{fig cycle43}--\ref{fig netzw}). 
It is well understood how  a steady 
state equilibrium is established in hot CNO-like reaction chains 
at temperatures $T_{9}\simeq 0.2$  and on  time 
scales governed by the longest $\beta^{+}$-decay in the cycle. 
Then the abundance $Y_{i,j}$ of nucleus ($i$)
in a cycle ($j$) can be calculated from the equilibrium condition
%(`` ! '' means ``condition'')

\begin{equation}
\dot{Y}_{i,j}\stackrel{!}{=}0.
\label{eq einlpa1c}
\end{equation}

This condition  can be 
used for simplifying the system of equations  which determines the 
thermonuclear reaction flows. However, an approximation of this kind 
is  no longer justified, when we consider time scales shorter than 
the half-lives of the $\beta^{+}$-unstable nuclei or 
temperatures greater than
$T_{9}$=$1$ where photodisintegrations become 
so important that 
a $(p,\gamma)$--$(\gamma,p)$-equilibrium is attained rather 
than a steady flow (see Figure \ref{fig netzw}). 
Then, proton-induced reactions
can be associated with  large quasi-decay constants
and, in comparison to $\beta$-decays, small half-lives.
Thus, all the abundances of isotonic nuclei which are connected by proton 
capture reactions can be considered in a quasistatic equilibrium
on time scales exceeding the half-lives associated with the quasi-decay 
constants. Such sequences shall be now defined as quasi-nuclei.

In the
following paragraphs, the abundances of nuclei in  proton capture
sequences shall be calculated under the assumption of a quasi-static
equilibrium.
Consider the  $j^{th}$ of $m$ sequences which consists of 
$n$$-$$1$ proton captures between $n$ nuclei ({\em cf.}  Figure 
\ref{fig diagr2}: Here, two proton capture sequences are
shown with $n$=$4$ and $n$=$5$). 
Each nucleus ($i$) in a sequence ($j$) is involved in 
$k+2$  reactions including  proton captures and photodisintegrations. 
Furthermore, we assume for simplicity that the abundances 
of helium and  hydrogen stay constant over the time scale of interest
and that nucleus $(i,j)$ is not fed by reactions from another chain $j'$.
Using the equilibrium condition (\ref{eq einlpa1c}),
the  temporal change of the abundance of a nucleus $(i,j)$ can 
be written as

\begin{eqnarray}
\dot{Y}_{i,j}&=&\lambda^{p,\gamma}_{i-1,j}Y_{i-1,j}-\sum_{k}\lambda^{x}_{i,j,k}
Y_{i,j} \nonumber \\
&&+\lambda^{\gamma,p}_{i+1,j} Y_{i+1,j}=0\; .
\label{eq einlpa1}
\end{eqnarray}

\noindent
The terms summed over $k$ are leaks out of chain ($j$) via nucleus $(i,j)$
The extreme cases of flux equilibria, like steady flows on the one hand
$(\lambda^{p,\gamma}\gg \lambda^{\gamma,p},\,\,\lambda^{\beta^{+}})$ and
$(p,\gamma)$--$(\gamma,p)$-equilibria on the other hand
$(\lambda^{p,\gamma},\,\, \lambda^{\gamma,p} \gg \lambda^{\beta^{+}})$
can be discussed
on the basis of this equation. 
Here, the general case shall be discussed.

For a chain of $n$ nuclei, $n$$-$$1$ such equations 
permit to relate the abundance
$Y_{i-1,j}$ to $Y_{i,j}$ starting with a ``known'' $Y_{n,j}$, the abundance 
of the last nucleus in a sequence.
Thus we obtain for each sequence ($j$) an $(n$$-$$1)$-dimensional 
system of linear equations
which allows us to express each abundance $Y_{i,j}$  in terms of
$Y_{n,j}$:

\begin{equation}
Y_{i,j}=v_{i,j}(\lambda^{x}_{1,j}\dots\lambda^{x}_{n,j}) Y_{n,j}\; .
\label{eq einlpa1b}
\end{equation}

\noindent
The $v_{i,j}$ depend only on the  constant or slowly varying quasi-decay 
constants $\lambda_{i,j}$.
The  abundance of a sequence can be treated as the abundance of
a quasi-nucleus by
$Y_{j}$:=$\sum_{i} Y_{i,j}$. When $Y_{j}$ is already
determined at a point in time $t$=$t_{s}$, each abundance $Y_{i,j}$
can be written as

\begin{equation}
Y_{i,j}(t_{s})=\frac{v_{i,j}}{v_{j}}Y_{j}(t_{s})\; ,
\label{eq einlpa2}
\end{equation}

\noindent
with $v_{j}$=$\sum_{i}v_{i,j}$. Hence, 
when the abundances of the quasi-nuclei $Y_{j}$ are
given at a time $t$, the abundances $Y_{i,j}(t)$ in each sequence 
can be  determined {\em simultaneously}.

\alphfg
\alphfg
%\clearpage
%\thispagestyle{empty}
\begin{figure}[tb]
\epsscale{0.75}
\figurenum{7a}
\plotfiddle{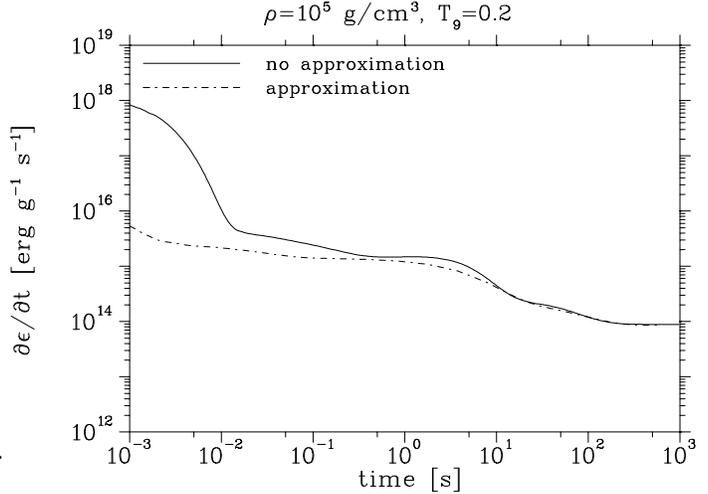}{5.5 cm}{90}{40}{40}{130}{-50}
\figcaption[eps52p1.ps]{Comparison of the energy generation rate 
$\dot{\epsilon}$ for for nova-like conditions.\label{fig novae1}}
\end{figure}

%\clearpage
%\thispagestyle{empty}
\begin{figure}[tb]
\epsscale{0.75}
\figurenum{7b}
\plotfiddle{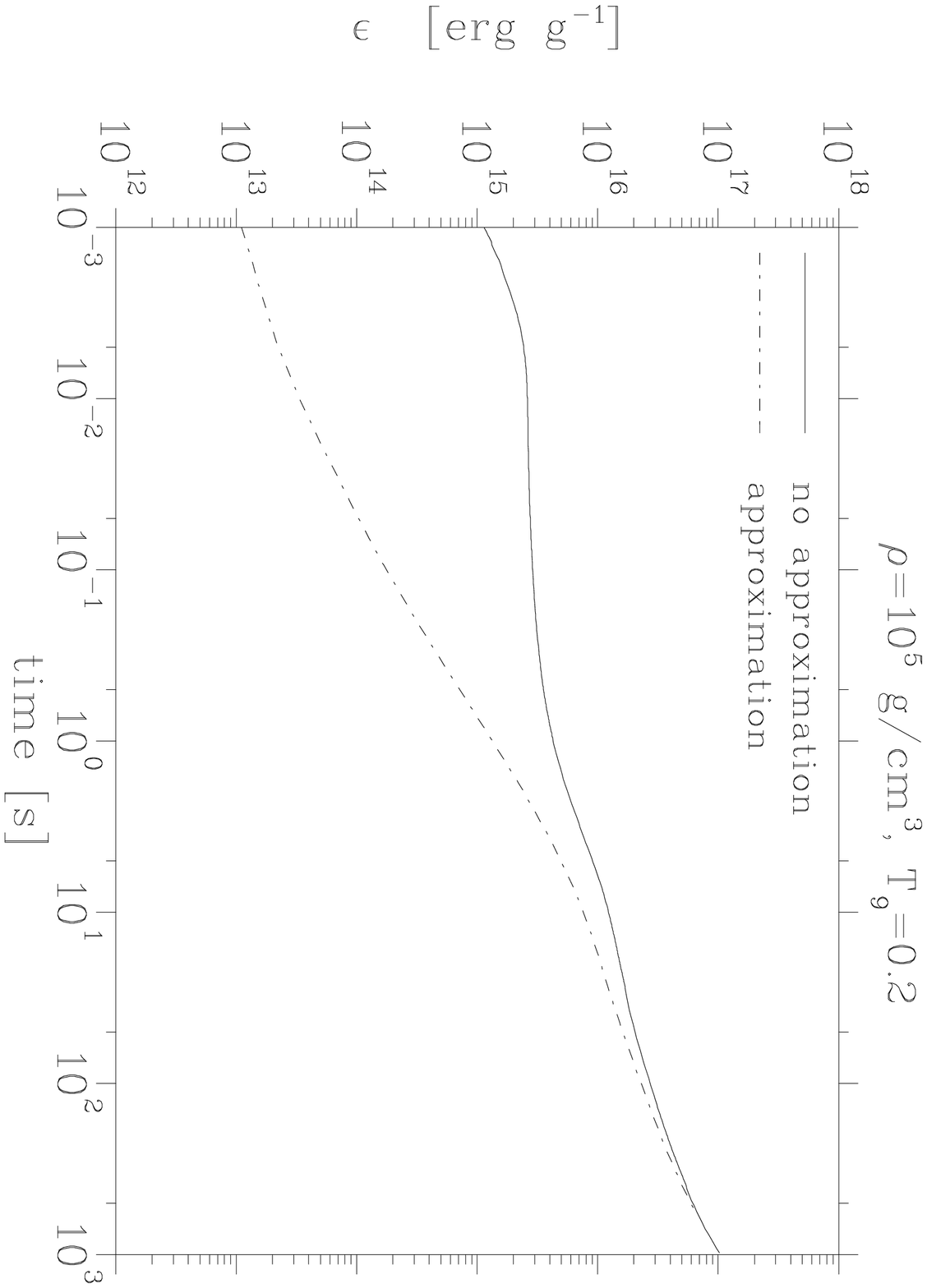}{5.5 cm}{90}{40}{40}{130}{-50}
\figcaption[ener52p1.ps]{Comparison of the energy generation 
$\epsilon$ for nova-like conditions.\label{fig novae2}}
\end{figure}

%\clearpage
%\thispagestyle{empty}
\begin{figure}[htb]
\figurenum{7c}
\epsscale{0.75}
\plotfiddle{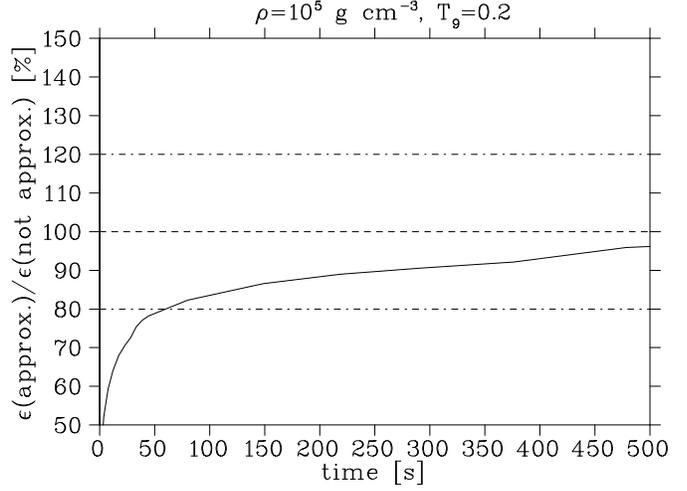}{5.5 cm}{90}{40}{40}{130}{-50}
\figcaption[epscheck52p1.ps]{With this
approximation, energy generation for nova-like conditions 
can be reproduced with an accuracy 
of 15-5 per cent.\label{fig novae3}}
\end{figure}

However, we also have to take care of incoming fluxes into a sequence 
$(j)$ not jet considered in equation (\ref{eq einlpa1}). As a
consequence, the coefficients $v_{i,j}$
are no longer constant because they 
become explicitly  abundance-dependent  
for the following reason:
in contrast to the {\em outgoing} fluxes,
the {\em incoming} fluxes into a sequence $j$ are 
{\em not} proportional to the abundances $Y_{i,j}$
of the nuclei in this sequence. 
Therefore,  abundances
can no longer be determined simultaneously.

Incoming fluxes can  nevertheless be considered 
as long as the abundances do not change too fast during a time step.
We can add to equation (\ref{eq einlpa1}) all incoming
fluxes  from nuclei  ($l$) of different chains ($m$)  proportional to
their abundances $Y_{l,m}(t_{s-1}),\,\, m$$\not=$j (when no
time $t$ is indicated, the quantities are given at  $t$=$t_{s}$),

%\samepage{
\begin{eqnarray}
\dot{Y}_{i,j}&=&-Y_{i,j}\sum_{k}\lambda^{x}_{i,j,k}+\lambda^{\gamma,p}_{i+1} 
Y_{i+1,j}+\lambda^{p,\gamma}_{i-1,j}Y_{i-1,j}\nonumber\\[0.3 cm]
& & +\sum_{l,m \atop m\neq j} \lambda^{x}_{l,m} Y_{l,m}(t_{s-1})=0\; .
\label{eq einlpa1d}
\end{eqnarray}

\noindent
This can be solved for $Y_{i-1,j}$, leading to

\begin{eqnarray}
Y_{i-1,j}&=&\underbrace{
\frac{1}{\lambda^{p,\gamma}_{i-1,j}}\left(Y_{i,j}\sum_{k}\lambda^{x}_{i,j,k}
-\lambda^{\gamma,p}_{i+1}Y_{i+1,j}\right)}_{v_{i-1,j}\,Y_{n,j}}\nonumber\\[0.3cm]
&& -\underbrace{\frac{1}{\lambda^{p,\gamma}_{i-1,j}}\sum_{l,m \atop m\neq j} 
\lambda^{x}_{l,m} Y_{l,m}(t_{s-1})
}_{w_{i-1,j}(\lambda_{l,m}^{x},Y_{l,m}(t_{s-1}))}\, ,
\end{eqnarray}

\noindent
or more generally

\begin{equation}
Y_{i,j}=v_{i,j}\, Y_{n,j}-w_{i,j}(t_{s-1})\, .
\label{eq einlpa3}
\end{equation}

\noindent
When  a slowly varying function $v'_{j}(t_{s-1})$ is given so 
that we can approximate  $Y_{n,j}$  by
 
\begin{equation}
Y_{n,j}(t_{s})\approx \frac{1}{v'_{j}(t_{s-1})}Y_{j}(t_{s})\; ,
\label{eq einlpa4}
\end{equation}

\noindent
we can extend $w_{i,j}$ in (\ref{eq einlpa3}) as follows

\begin{equation}
Y_{i,j}=v_{i,j}Y_{n,j}-\frac{v'_{j}}{Y_{j}}w_{i,j}Y_{n,j}\, ,
\end{equation}

\noindent
and finally obtain

%\begin{equation}
%\fbox{$ \displaystyle Y_{i,j} = v'_{i,j} Y_{n,j} $}
%\label{eq einlpa 5}
%\end{equation}

\begin{equation}
 Y_{i,j} = v'_{i,j} Y_{n,j} 
\label{eq einlpa 5}
\end{equation}

\begin{equation}
v'_{i,j}=v_{i,j}-\frac{v'_{j}}{Y_{j}}w_{i,j}\,  ,
\end{equation}

\noindent
with $v'_{j}$=$\sum_{i}v'_{i,j}$. Setting 
$v_{i,j}'(t_{0}$=$0)$=$v_{i,j}(t_{0}$=$0)$ as
initial values, the existence of the functions $v'_{j}$ is guaranteed. 

So far we have made two approximations. (a): including reactions within
a chain and leaks out of a chain described by equation (\ref{eq einlpa1}).
In that case all abundances could be described simultaneously. 
(b): When also considering incoming fluxes [see equation (\ref{eq einlpa1d})], 
we become
dependent on abundances in other chains. Approximating their abundances at 
$t$=$t_{s}$ by abundances at $t'$=$t_{s-1}$ permits to stay within the same 
approximation scheme. But we have to be aware that this is only valid for 
slowly varying abundances or equivalently small time steps.

\begin{figure}[tb]
\epsscale{0.75}
\figurenum{8a}
\plotfiddle{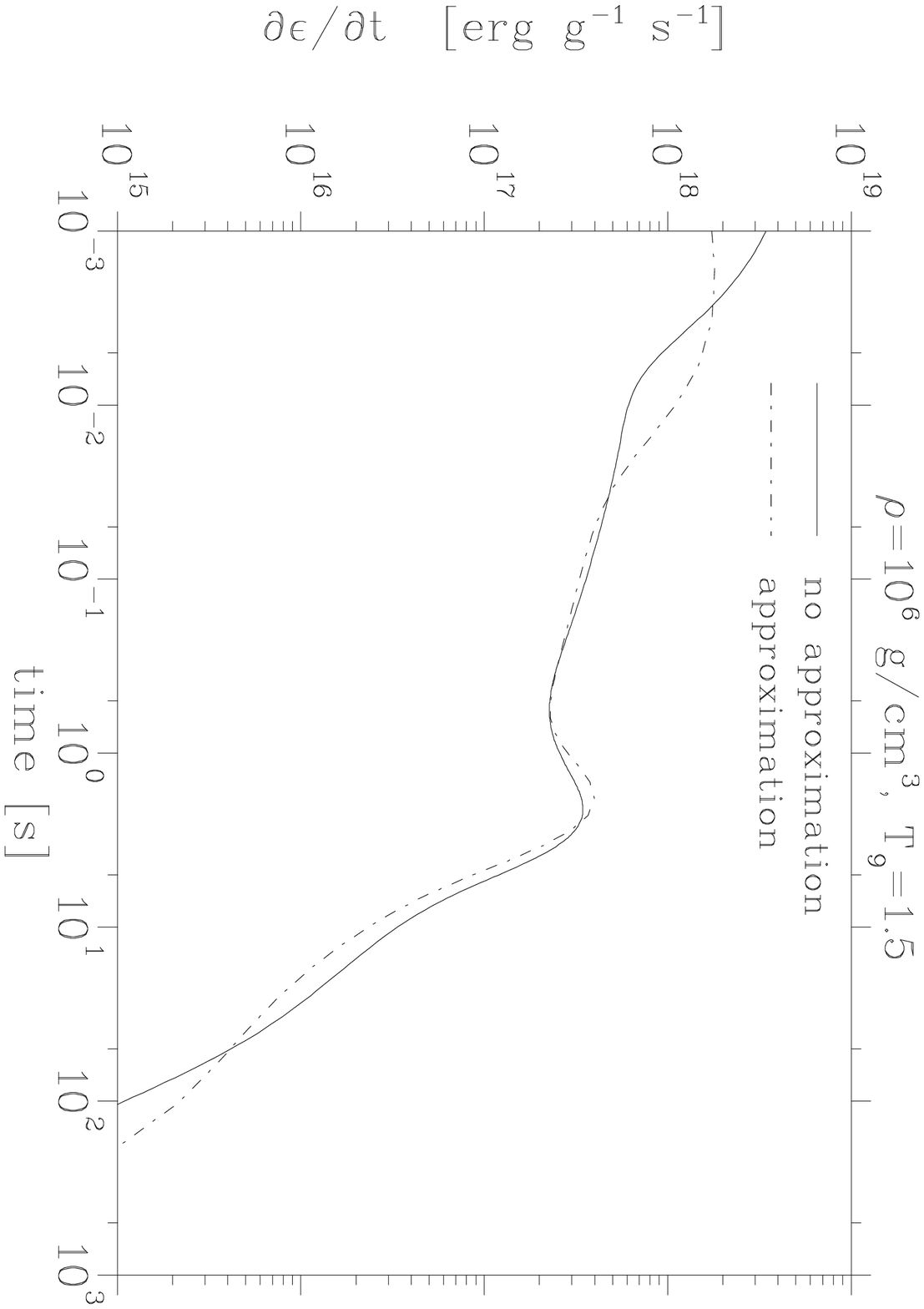}{5.5 cm}{90}{40}{40}{130}{-50}
\figcaption[eps615p1.ps]{Same as Figure \ref{fig novae1}
but for X-ray burst conditions.\label{fig xrayb1}}
\end{figure}

\begin{figure}[tb]
\epsscale{0.75}
\figurenum{8b}
\plotfiddle{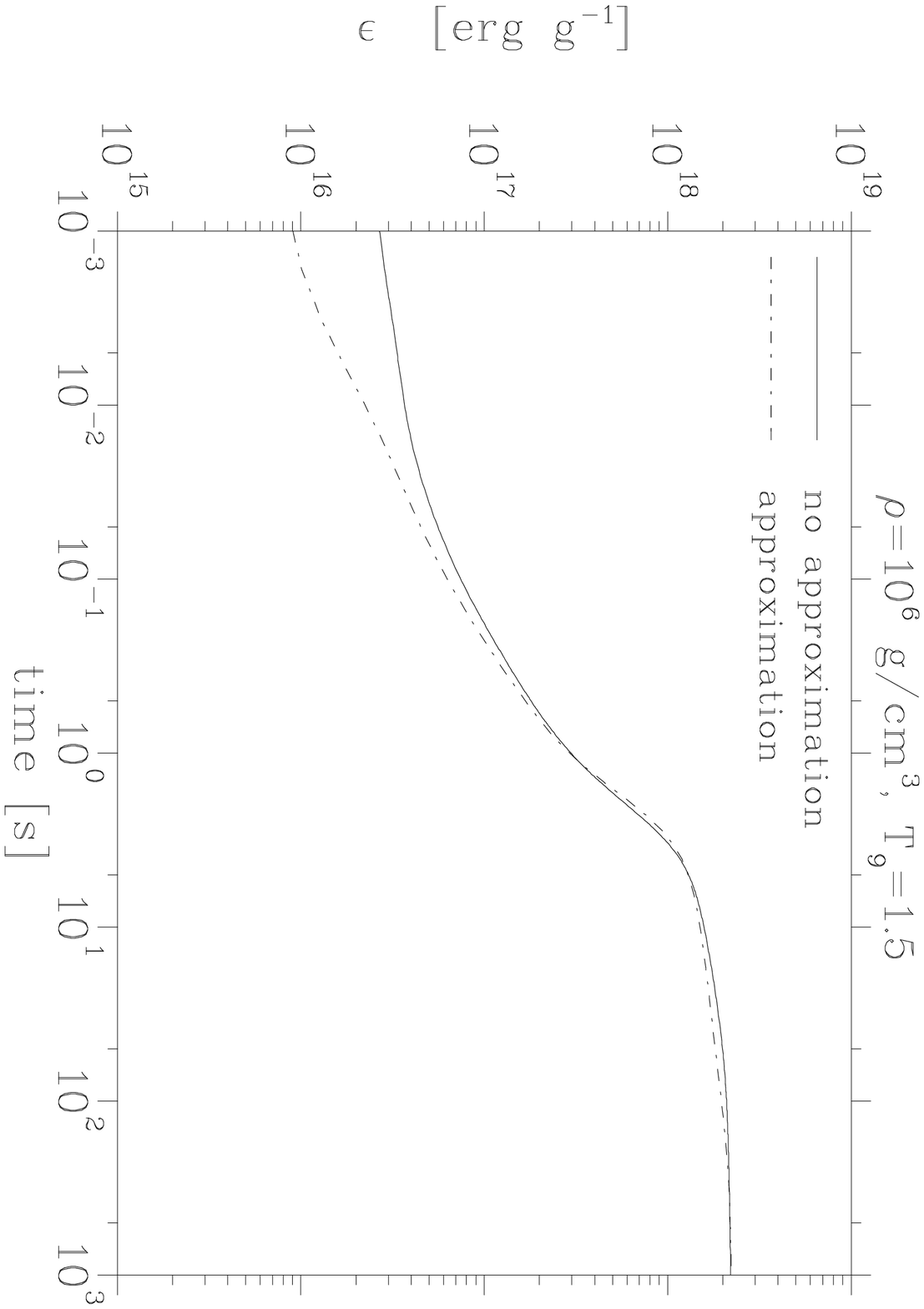}{5.5 cm}{90}{40}{40}{130}{-50}
\figcaption[ener615p1.ps]{Same as Figure \ref{fig novae2}
but for X-ray burst conditions.\label{fig xrayb2}}
\end{figure}

\begin{figure}[tb]
\epsscale{0.75}
\figurenum{8c}
\plotfiddle{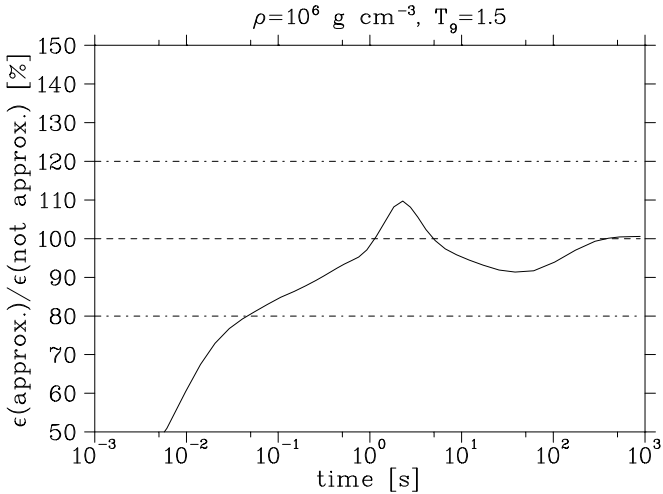}{5.5 cm}{90}{40}{40}{130}{-50}
\figcaption[epscheck615p1.ps]{Same as Figure \ref{fig novae3}
but for X-ray burst conditions.\label{fig xrayb3}}
\end{figure}

\placefigure{fig xrayb1}
\placefigure{fig xrayb2}
\placefigure{fig xrayb3}
%\placefigure{fig intcon1}
%\placefigure{fig intcon2}
%\placefigure{fig intcon3}
%\newpage

\resetfg
\alphfg
%\clearpage
%\thispagestyle{empty}

%\clearpage
%\thispagestyle{empty}

%\thispagestyle{empty}
%\clearpage
\resetfg

%Let us discuss the two approximations made so far: From a
%``physical'' point of view, there is no reason to neglect incoming
%fluxes in the equilibrium condition when outgoing fluxes are taken into 
%account. But there is a mathematical, numerical one: The second
%approximation is numerically more unstable than the first one because
%quantities must be used at each time $t$=$t_{s}$ which were calculated 
%at a former time $t'$=$t_{s-1}$

We obtained this approximation under the assumption
that the abundances of hydrogen  and helium stay constant.
$Y_{p}$ and $Y_{\alpha}$ are slowly varying functions
in realistic environments.
%when temperature and density reach values where X-ray bursts
%may occur ($T_{9}$$\approx$$1$, $\rho$=$10^{6}$ g/cm$^{3}$) 
%due to the ongoing 
%proton- and alpha-induced reactions. 
%Especially $Y_{\alpha}$ changes significantly because of the
%$3\alpha$-reaction
%$^{4}$He(2$\alpha,\gamma)^{12}$C. 
We can adjust for the change of $Y_{p}$ and $Y_{\alpha}$ and
obtain their new values applying relation (\ref{eq einlpa2})

\begin{mathletters}
\begin{eqnarray}
\dot{Y}_{p}&=&\sum\limits^{}_{i,j} (\lambda^{\gamma,p}_{i,j}
+\lambda^{\alpha,p}_{i,j}-\lambda^{p,\alpha}_{i,j} \nonumber \\
&&-\lambda^{p,\gamma}_{i,j})Y_{i,j}(t_{s})\\
\dot{Y}_{\alpha}&=&\sum\limits^{}_{i,j}(\lambda^{p,\alpha}_{i,j}
 -\lambda^{\alpha,p}_{i,j}+\lambda^{\gamma,\alpha}_{i,j} \nonumber \\
&& -\lambda^{\alpha,\gamma}_{i,j}) Y_{i,j}\; .
\end{eqnarray}
\end{mathletters}

As a consequence, the quasi-decay constants must be recalculated
after each time step, making use of the updated proton and
alpha  abundances
considering them constant for one  time step.

Thus it is possible to express simultaneously 
every abundance $Y_{i,j}$ in a sequence
with the sequence abundance $Y_{j}$. With an
approximation scheme of this kind, we
reduced our full network described by $\sum_{j=1}^{m} n(j)$ equations
to a system of $m$ equations. In the following we want to check 
the accuracy of such an approximation.

\resetfg
%\thispagestyle{empty}
%\clearpage
\begin{figure}[htb]
\epsscale{0.75}
\figurenum{9}
\plotfiddle{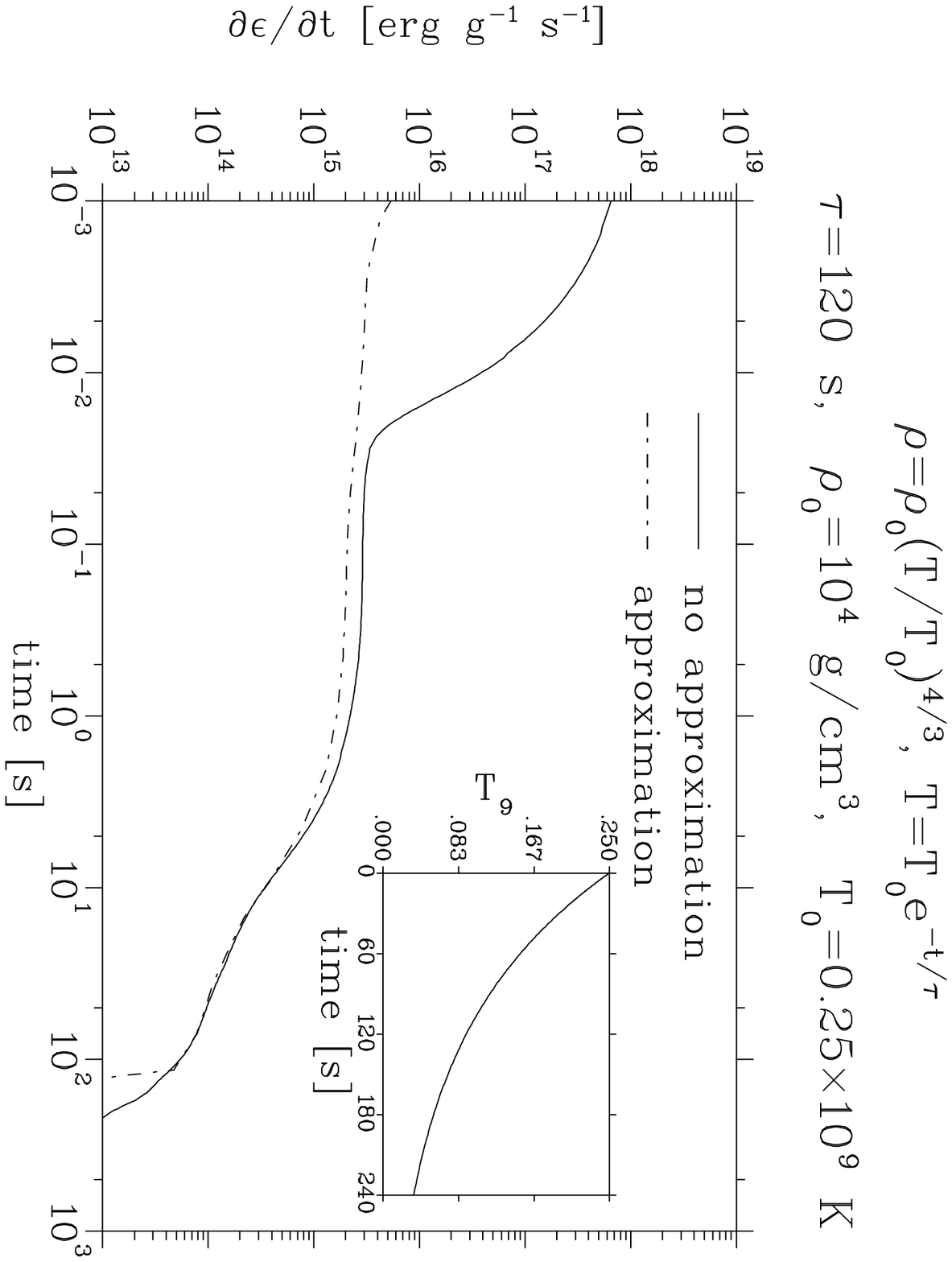}{5 cm}{90}{40}{40}{150}{-50}
\figcaption[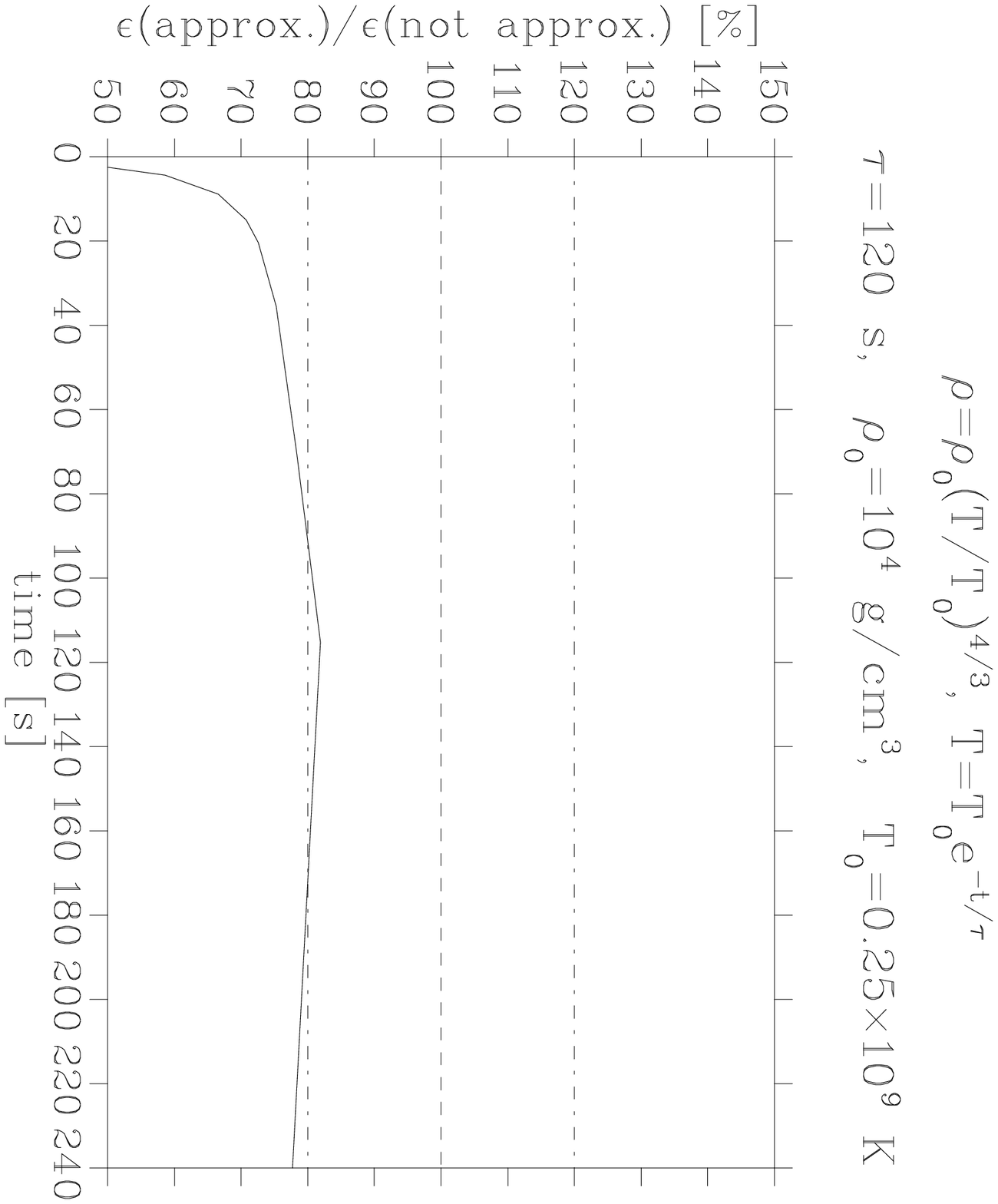]{Same as Figure \ref{fig novae1}
but for varying density and temperature 
starting with novae peak conditions. 
The approximation is valid as long as the
temperature does not fall considerably below $T_{9}=0.1$.
\label{fig novaevar1}}
\end{figure}

%\thispagestyle{empty}
%\clearpage
%\begin{figure}[htb]
%\epsscale{0.75}
%\figurenum{9b}
%\plotfiddle{enercheck1var52p1.ps}{5 cm}{90}{40}{40}{150}{-50}
%\figcaption[eps1var52p1.ps]{Same as Figure \ref{fig novae3}
%but for varying density and temperature 
%starting with novae peak conditions. \label{fig novaevar2}}
%\end{figure}

%\thispagestyle{empty}
%\clearpage
\begin{figure}[htb]
\epsscale{0.75}
\figurenum{10}
\plotfiddle{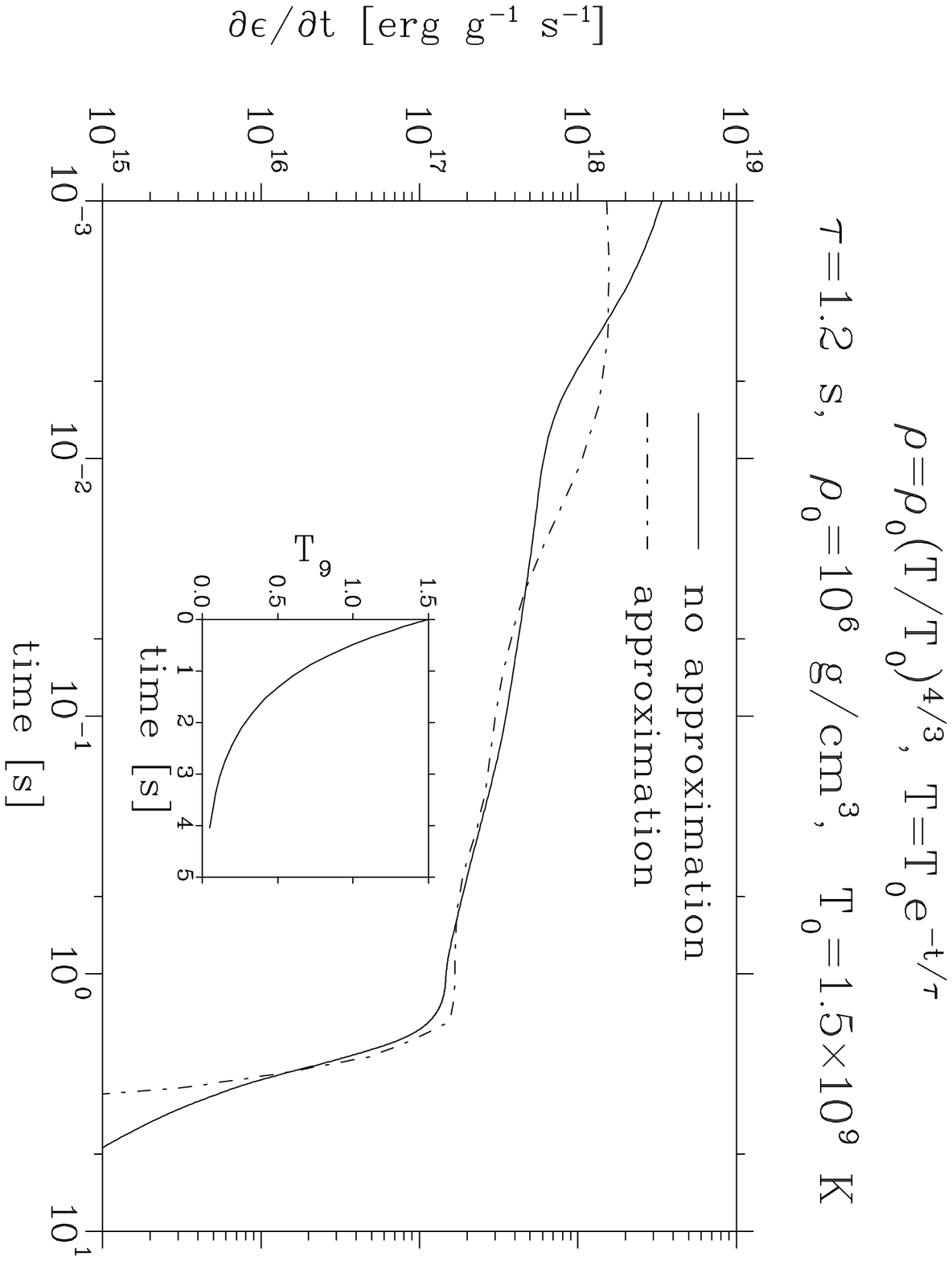}{5 cm}{90}{40}{40}{150}{-50}
\figcaption[eps1var615p1.ps]{Same as Figure \ref{fig novaevar1}
but for X-ray burst peak condition.
\label{fig xraybvar1}}
\end{figure}

%\thispagestyle{empty}
%\clearpage
%\begin{figure}[htb]
%\epsscale{0.75}
%\figurenum{10b}
%\plotfiddle{enercheck1var615p1.ps}{5 cm}{90}{40}{40}{150}{-50}
%\figcaption[enercheck1var615p1.ps]{Same as Figure \ref{fig novae3}
%but for varying density and temperature
%starting with X-ray burst peak conditions.\label{fig xraybvar2}}
%\end{figure}

\section{RESULTS}\label{results}
The calculations were done with a reaction network which consists 
mainly of proton capture sequences, each starting out from a
$T_{z}$=$0$ nucleus. The number of nuclei in a sequence is determined by the 
condition that the quasi decay constant ${^{i+1}\lambda}_{i}^{p,\gamma}$
of a nucleus $(i)$ into nucleus ($i$+$1$) is larger than a certain 
$\lambda_{\mbox{\tiny min}}$. When this condition is fulfilled, nucleus ($i$+$1$) 
is considered to be in the sequence.
For the following calculations $\lambda_{\mbox{\tiny min}}$ was set to $10^{3}$~s$^{-1}$ 
because we wanted to reproduce the energy generation of explosive 
hydrogen burning on time scales
$\tau$=$\lambda_{\mbox{\tiny min}}^{-1}>10^{-2}-10^{-1}$ s. As a consequence, a 
sequence contains  not more than 5 nuclei for
temperatures $T_{9}$$\leq$$1.5$ and densities $\rho$$\leq$$10^{6}$~g~cm$^{-3}$.

In addition to the proton capture sequences, 
$T_{z}$=$\frac{1}{2}$ nuclei had to be considered in the 
network. They are the most neutron-rich nuclei of the hot CNO-type cycles,
where after a proton capture, the alpha-threshold in the compound nucleus 
is lower than the proton-threshold, and $(p,\alpha)$-reactions can occur
which compete with the still possible $(p,\gamma)$-transitions 
(see Figure \ref{fig diagr2}). 
Since the $(p,\alpha)$- and $(p,\gamma)$-reactions
occur on the same time scale, these $T_{z}$=$\frac{1}{2}$ nuclei
can not affiliated to a proton capture chain but have to
be considered separately.
 
Between $^{12}$C and $^{50}$Mn, the 
network can be composed of proton capture reaction sequences and 
$T_{z}$=$\frac{1}{2}$ nuclei with one exception. 
The transition occurs  already at compound nucleus
$^{19}$Ne in the Ne isotopic chain. A $(p,\alpha)$-reaction 
is possible on the $T_{z}$=$0$
nucleus $^{18}$F and a proton capture reaction sequence starts out
from $^{19}$Ne with isospin  $T_{z}$=$-\frac{1}{2}$.

The low alpha-threshold of $^{19}$Ne,
which caused an exception of the rule and prevented an OFNe-cycle, permits
on the other hand the $^{15}$O($\alpha,\gamma)$-reaction and triggers
the break-out from the CNO-cycles. 
The network described so far is shown in the Figure \ref{fig diagr1} and
\ref{fig diagr2}.

At temperatures exceeding $T_{9}$=$0.8$ and $\rho$ $\geq$ $10^{6}$~g~cm$^{-3}$,
the network must be completed with the $3\alpha$-reaction.
Furthermore, it is very likely that fluxes up to $^{72}$Kr
may influence the energy generation. 
In the region between $^{51}$Mn and $^{72}$Kr, the
$\beta^{+}$-reactions of the for the rp-process relevant
nuclei are, with the exception of $^{55}$Ni, $^{64}$Ge and $^{72}$Kr,
considerably smaller than the typical time scales
of novae and X-ray bursts. Consequently, only
the above mentioned nuclei were considered explicitly
in the approximation scheme. 
The approximated abundances of $^{53}$Ni and these three waiting point nuclei 
are compared with full network 
calculation results in Figure \ref{fig abpa1} and \ref{fig abpa2}.

The nuclei used in the approximation scheme
are listed in Table 1 according to their affiliation to a
certain proton capture reaction chain. Nuclei which had to
be considered separately (p, $^{4}$He, $T_{z}$=$\frac{1}{2}$ nuclei,
$^{55}$Ni, $^{64}$Ge and $^{72}$Kr) are given an own sequence number.

\noindent
The waiting point abundances 
obtained with this approximation are compared with full network calculation
results in Figure \ref{fig abpa1} and \ref{fig abpa2}.
  
Approximating the network according to section \ref{approximation},
we calculated the energy generation rate $\dot \epsilon$ 
and the integrated energy generation $\epsilon$ 
for nova-like conditions
($T_{9}$=0.2, $\rho$=$10^{5}$~g~cm$^{-3}$)  and X-ray burst peak conditions
($T_{9}$=1.5, $\rho$=$10^{6}$~g~cm$^{-3}$)
and compared the results to full size network calculations which
take into account each nucleus separately. 
In each case we assumed a solar abundance distribution for the
accreted matter.
The results for small temperatures 
are shown in Figure \ref{fig novae1}--\ref{fig novae3} 
and those for X-ray burst conditions 
in Figure \ref{fig xrayb1}--\ref{fig xrayb3}.
Considering the time scales important for novae (100--1000~s) and X-ray bursts
(1--10~s), we see that the approximation scheme devised in section 3 agrees
always to a high degree for the astrophysical applications considered here. 
The overall results show only deviations smaller than 15 per cent or even
less (down to 5 per cent) at a gain in computational speed by
a factor of 15. Several calculations at intermediate 
temperatures and densities, i. e. $10^{4}$$\le$$\rho$$\le$$10^{6}$,
$0.2$$\le$$T_{9}$$\le$$1.5$
confirmed  the above mentioned efficiency and accuracy
of the approximation scheme.

\placefigure{fig novae1}
\placefigure{fig novae2}
\placefigure{fig novae3}

Such an approximation  would not be very useful if
it was only valid for constant temperatures and densities.
In order to check its accuracy
for varying temperatures and densities we assumed 
after peak conditions were obtained in the ignition
an exponentially decreasing temperature 
and an adiabatically expanding accretion layer
with a radiation dominated equation of state ($\gamma=\frac{4}{3}$):

 \begin{mathletters}
\begin{eqnarray}
T&=&T_{0}e^{-\frac{t}{\tau}}\\
\rho&=&\rho_{0}\left(\frac{T}{T_{0}}\right)^{\frac{4}{3}}.
\end{eqnarray}
\end{mathletters}

Using typical peak temperatures $T_{0}$, densities $\rho_{0}$
and time scales $\tau$ for X-ray bursts and novae, we again calculated
the energy generation (Figure \ref{fig novaevar1} and \ref{fig xraybvar1}). 
The results show that the approximation is valid for changing
temperatures and densities as long as  they are in the range of
$10^{4}$$\le$$\rho$$\le$$10^{6}$, $0.1$$\le$$T_{9}$$\le$$1.5$. No
significant change in computational speed was observed.

\placefigure{fig novaevar1}
\placefigure{fig novaevar2}
\placefigure{fig xraybvar1}
\placefigure{fig xraybvar2}
%\nopagebreak
The differences at  time scales 
smaller than $t$=$10^{-2}$~s
are due to the very fast reaction flows within the proton capture 
reaction chains: They  occur in full scale network calculations  
until equilibrium 
is reached according to equation  (\ref{eq einlpa1c}).

%\begin{table}[h]
\begin{center}
{\sc Table 1:\\ Proton Capture Sequences$^{1}$\/}
\begin{tabular}{p{3.5 cm} r}
%\tablenum{1}
\small
\tablewidth{6 cm}
%\tablehead{\colhead{Sequence Number} & \colhead{Nuclei}} 
%\startdata
&\\
\hline
\hline
Sequence Number & Nuclei \\
\hline
Sequence 1 \dotfill & p  \\
%\tablevspace{3 mm}
Sequence 2 \dotfill & $^{4}$He  \\
%\tablevspace{3 mm}
Sequence 3 \dotfill & $^{12}$C \\
& $^{13}$N \\
& $^{14}$O \\
%\tablevspace{3 mm}
Sequence 4 \dotfill & $^{14}$N \\
& $^{15}$O \\
%\tablevspace{3 mm}
Sequence 5 \dotfill & $^{15}$N \\
%\tablevspace{3 mm}
\end{tabular}
\end{center}

\begin{center}
{\sc Table 1 --- \em Continued \/}
\begin{tabular}{p{3.5 cm} r}
& \\
\hline
\hline
Sequence Number & Nuclei \\
\hline
Sequence 6 \dotfill & $^{19}$Ne \\
& $^{20}$Na\\
& $^{21}$Mg\\
%\tablevspace{3 mm}
Sequence 7 \dotfill & $^{16}$O \\
& $^{17}$F \\
& $^{18}$Ne \\
%\tablevspace{3 mm}
Sequence 8 \dotfill & $^{20}$Ne \\
& $^{21}$Na \\
& $^{22}$Mg \\
& $^{23}$Al \\
& $^{24}$Si \\
%\tablevspace{3 mm}
%\tablebreak
Sequence 9 \dotfill & $^{22}$Na \\
& $^{23}$Mg \\
& $^{24}$Al \\
& $^{25}$Si \\
%\tablevspace{3 mm}
%\end{tabular}
%\end{center}
%\end{table}
Sequence 10 \dotfill & $^{23}$Na \\
%\tablevspace{3 mm}
Sequence 11 \dotfill& $^{24}$Mg \\
& $^{25}$Al \\
& $^{26}$Si \\
& $^{27}$P \\
& $^{28}$S \\
%\tablevspace{3 mm}
%\end{table}
Sequence 12 \dotfill & $^{26}$Al \\
& $^{27}$Si \\
& $^{28}$P \\
& $^{29}$S  \\
%\tablevspace{3 mm}
Sequence 13 \dotfill & $^{27}$Al \\
%\tablevspace{3 mm}
Sequence 14 \dotfill & $^{28}$Si \\
& $^{29}$P \\
& $^{30}$S \\
& $^{31}$Cl \\
& $^{32}$Ar \\
%\tablevspace{3 mm}
%\tablebreak
Sequence 15 \dotfill & $^{30}$P \\
& $^{31}$S \\
& $^{32}$Cl \\
& $^{33}$Ar \\
%\tablevspace{3 mm}
Sequence 16 \dotfill & $^{31}$P \\
%\tablevspace{3 mm}
\end{tabular}
\end{center}
\pagebreak

\begin{center}
{\sc Table 1 --- \em Continued \/}
\begin{tabular}{p{3.5 cm} r}
 & \\
\hline
\hline
Sequence Number & Nuclei \\
\hline
Sequence 17 \dotfill & $^{32}$S \\
& $^{33}$Cl \\
& $^{34}$Ar \\
& $^{35}$K \\
& $^{36}$Ca \\
%\tablevspace{3 mm}
Sequence 18 \dotfill & $^{34}$Cl \\
& $^{35}$Ar \\
& $^{36}$K \\
& $^{37}$Ca \\
%\tablevspace{3 mm}
Sequence 19 \dotfill & $^{35}$Cl \\
%\tablevspace{3 mm}
Sequence 20 \dotfill & $^{36}$Ar \\
& $^{37}$K \\
& $^{38}$Ca \\
& $^{39}$Sc \\
& $^{40}$Ti \\
%\tablevspace{3 mm}
Sequence 21 \dotfill & $^{38}$K \\
& $^{39}$Ca \\
& $^{40}$Sc \\ 
& $^{41}$Ti \\
%\tablevspace{3 mm}
Sequence 22 \dotfill & $^{39}$K \\
%\tablebreak
%\tablevspace{3 mm}
Sequence 23 \dotfill & $^{40}$Ca \\
& $^{41}$Sc \\
& $^{42}$Ti \\
& $^{43}$V \\
& $^{44}$Cr \\
%\tablevspace{3 mm}
Sequence 24 \dotfill & $^{42}$Sc \\
& $^{43}$Ti \\
& $^{44}$V \\
& $^{45}$Cr \\
%\tablevspace{3 mm}
Sequence 25 \dotfill & $^{43}$Sc \\
%\tablevspace{3 mm}
Sequence 26 \dotfill & $^{44}$Ti \\
& $^{45}$V \\
& $^{46}$Cr \\
& $^{47}$Mn \\
& $^{48}$Fe \\
%\tablevspace{3 mm}
%\tablebreak
\end{tabular}
\end{center}
\pagebreak

\begin{center}
{\sc Table 1 --- \em Continued \/}
\begin{tabular}{p{3.5 cm} r}
 & \\
\hline
\hline
Sequence Number & Nuclei \\
\hline
Sequence 27 \dotfill & $^{46}$V \\
& $^{47}$Cr \\
& $^{48}$Mn \\
& $^{49}$Fe \\
%& $^{50}$Co \\ 
%\tablevspace{3 mm}
Sequence 28 \dotfill & $^{47}$V \\
%\tablevspace{3 mm}
Sequence 29 \dotfill & $^{48}$Cr \\
& $^{49}$Mn  \\
& $^{50}$Fe \\
& $^{51}$Co \\
& $^{52}$Ni \\
%\tablevspace{3 mm}
Sequence 30 \dotfill & $^{50}$Mn \\
& $^{51}$Fe \\
& $^{52}$Co \\
& $^{53}$Ni \\
%\tablevspace{3 mm}
%\tablevspace{3 mm}
Sequence 31 \dotfill & $^{51}$Mn \\
%\tablevspace{3 mm}
Sequence 32 \dotfill & $^{55}$Ni  \\
%\tablevspace{3 mm}
Sequence 33 \dotfill & $^{64}$Ge \\
%\tablevspace{3 mm}
Sequence 34 \dotfill & $^{72}$Kr \\
\hline
%\enddata
\end{tabular}
\begin{tabular}{p{7 cm}}
{\footnotesize
$^{1}$The nuclei used in the appro\-ximation scheme
are listed  according to their affiliation to a
certain proton capture reaction chain. Nuclei which had to
be considered separately (p, $^{4}$He, $T_{z}$=$\frac{1}{2}$ nuclei,
$^{55}$Ni, $^{64}$Ge and $^{72}$Kr) are given an own sequence
number.\/} 
\end{tabular}
\end{center}

\section{CONCLUSION}\label{conc}
The main motivation for the present investigation was to find a fast and
efficient approximation scheme which permits to predict the energy generation 
in explosive hydrogen burning for a large range of conditions, i.e.
temperatures  $(0.2$$\le$$T_{9}$$\le$$ 
1.5)$ and densities $(10^{4}$~g~cm$^{-3}$$\le$$\rho$$\le 10^{6}$~g~cm$^{-3}$) 
respectively, for applications in hydro calculations which cannot afford
full nuclear network. This goal has been achieved.
The energy generation of full network calculations 
(150 nuclei) could be reproduced with a high accuracy, resulting in
deviations of 5 to a maximum of 15 per cent while gaining a factor of about
15 in computational speed.

The approximation scheme discussed in this paper is therefore well suited 
for realistic hydro calculations of novae or X-ray bursts and other
possible sites of explosive hydrogen burning.

\begin{figure}[tbn]
\epsscale{0.75}
\figurenum{11}
\plotfiddle{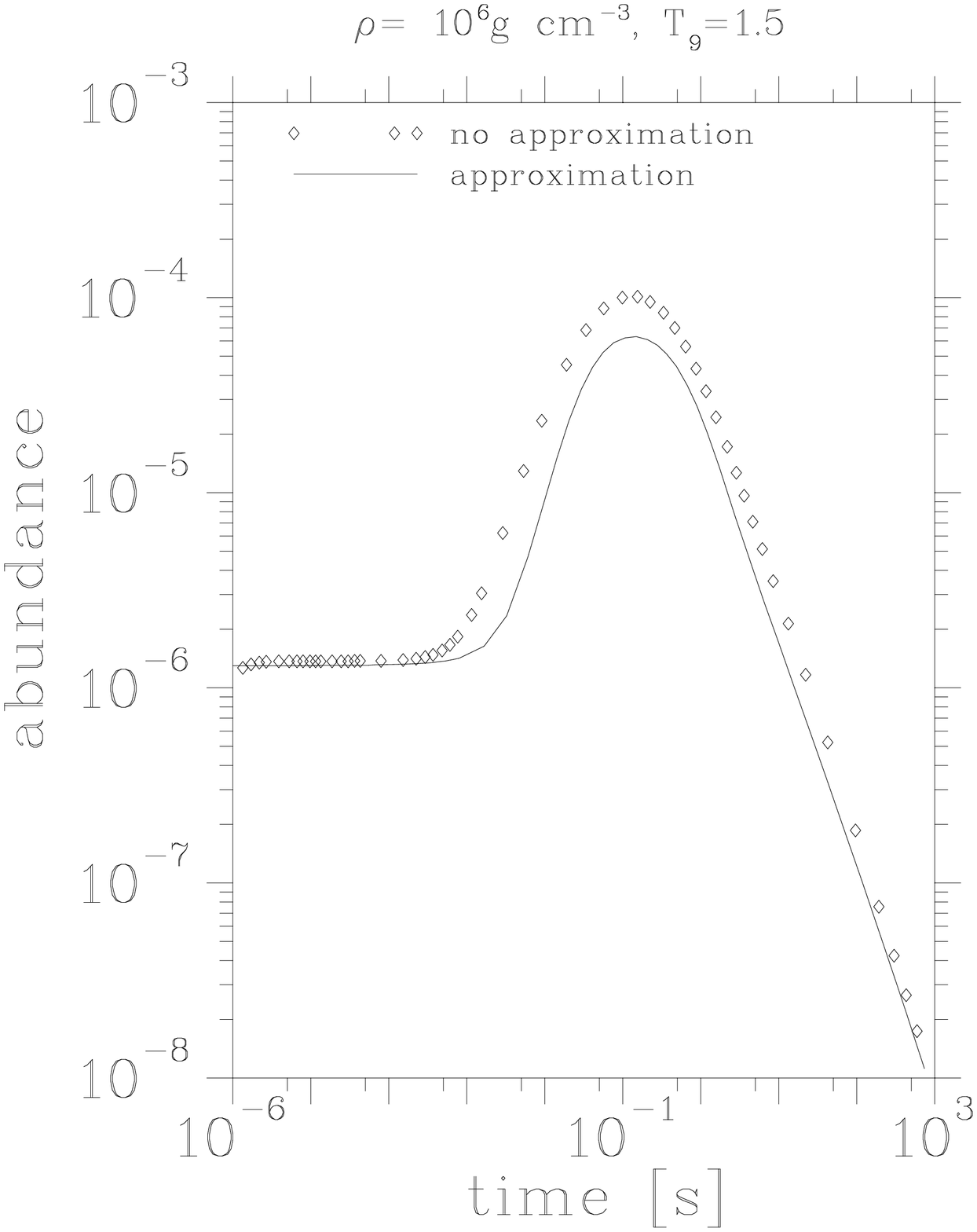}{6 cm}{0}{30}{30}{-100}{-20}
\figcaption[ar33615p1.ps]{Comparison of the  $^{33}$Ar-abundances
calculated by a  full network calculation and
the approximation scheme. \label{fig ar33615}}
\end{figure}

The additional advantage of this method is that it also permits to predict
isotopic abundances with a similar accuracy and can thus even replace
full network calculations for this purpose (see Figure \ref{fig ar33615}).

\resetfg
%\clearpage
%\thispagestyle{empty}

We were also able to determine key nuclear properties, which 
directly enter the precision of calculations for explosive hydrogen
burning, its energy generation and abundance determination. 
We have identified the  even-$Z$ T$_z$=--1/2 nuclei like $^{23}$Mg, 
$^{27}$Si, $^{31}$S, $^{35}$Ar and $^{39}$Ca, as essential targets for 
proton captures which are in competition with $\beta$-decays. As their
small reaction Q-values (less than 2~MeV) do not permit the application
of statistical model cross sections, they have to be determined experimentally.
Thus, these results are an ideal guidance 
for future experimental investigations with radioactive ion beams.
Nuclei closer to stability permit the application of statistical model
cross sections. Nuclei more proton-rich than $T_z$=-1 are only populated
for conditions when a $(p,\gamma)$--$(\gamma,p)$-equilibrium is established.

Thus, only their masses, spins and half-lives enter the calculation. Connecting
$(\alpha,p)$-reactions have high enough Q-values, even close to the
drip-line, that the level densities are sufficient for statistical model
cross sections (this is true at least up to Ca and Ti, where alpha-induced
reactions can play an important role). 
\nopagebreak
Therefore, the remaining experimental properties which should be determined
and enter the calculations in a crucial way are $\beta^+$-decay
half-lives and nuclear masses.  The latter control the abundances in a
$(p,\gamma)$--$(\gamma,p)$-equilibrium.
Beyond Se, the details of the proton drip-line
are actually not that well known yet, but can influence the
endpoint of the rp-process. First calculations which include 2p-capture
reactions (Schatz et al. 1996)
indicate that it seems possible to produce nuclei with A=90-100 in
the rp-process under X-ray burst conditions, 
depending on the choice of mass formulae and 
half-life predictions
(similar to r-process studies). Especially a region of strong 
deformation and small $\alpha$-capture Q-values around $A$=80 seems to be
predicted with large variations among different mass models, and the even-even
$N$=$Z$ nuclei between A=68 and 100 play a dominant role.
\nopagebreak
\acknowledgements
The work of F.~R., Ch.~F., and F.-K.~T
was supported by the Swiss National Science Foundation grant
20-47252.96. 
T.~R. acknowledges support by an APART fellowship
from the Austrian Academy of Sciences.
H.~S. was supported by the German Academic Exchange Service (DAAD)
with a ``Doktorandenstipendium aus Mitteln des zweiten 
Hochschulsonderprogramms''. M.~W. was supported by the DOE grant
DE-FG02-95-ER40934.
%------------------------------------------------------------------------------
%                            BIBLIOGRAPHY
%------------------------------------------------------------------------------

\newpage

%------------------------------------------------------------------------------
%                         DIAGRAMS AND PICTURES
%------------------------------------------------------------------------------

%\figcaption[eps57p1.ps]{Same as Figure \ref{fig novae1} but
%for intermediate conditions.
%\label{fig intcon1}}

%\figcaption[ener57p1.ps]{Same as Figure \ref{fig novae2}
%but for intermediate conditions.\label{fig intcon2}}

%\figcaption[epscheck57p1.ps]{Same as Figure \ref{fig novae3}
%but for intermediate conditions.\label{fig intcon3}}

%--------------------------------------------------------------------------
%figures
%--------------------------------------------------------------------------

\onecolumn
\clearpage

\clearpage
\thispagestyle{empty}

%\thispagestyle{empty}
%\begin{figure}[hbtn]
%\figurenum{8a}
%\plotone{eps57p1.ps}
%\end{figure}

%\thispagestyle{empty}
%\begin{figure}[hbtn]
%\figurenum{8b}
%\plotone{ener57p1.ps}
%\end{figure}

%\samepage{
%\thispagestyle{empty}
%\begin{figure}[htnp]
%\figurenum{8c}
%\plotone{epscheck57p1.ps}
%\end{figure}}

\end{document}